# Enhanced superfluid density on twin boundaries in $Ba(Fe_{1-x}Co_x)_2As_2$

B. Kalisky[1,2,*], J.R. Kirtley[1,2,3], J.G. Analytis[1,2,4], Jiun-Haw Chu[1,2,4], A. Vailionis[1,4], I.R. Fisher[1,2,4], K.A. Moler[1,2,4,5,*]

[1] *Geballe Laboratory for Advanced Materials, Stanford University, Stanford, California 94305-4045, USA*

[2] *Department of Applied Physics, Stanford University, Stanford, California 94305-4045, USA*

[3] *IBM Watson Research Center, Route 134 Yorktown Heights, NY 10598, USA*

[4] *Stanford Institute for Materials and Energy Sciences, SLAC National Accelerator Laboratory, 2575 Sand Hill Road, Menlo Park, CA 94025*

[5] *Department of Physics, Stanford University, Stanford, California 94305-4045, USA*

* authors to whom correspondence should be addressed: Beena Kalisky e-mail: beena@stanford.edu. K.A. Moler e-mail: kmoler@stanford.edu.

Abstract:
Superconducting quantum interference device (SQUID) microscopy shows stripes of increased diamagnetic susceptibility in underdoped, but not overdoped, single crystals of $Ba(Fe_{1-x}Co_x)_2As_2$. These stripes of increased diamagnetic susceptibility are consistent with enhanced superfluid density on twin boundaries. Individual vortices avoid pinning on or crossing the stripes, and prefer to travel parallel to them. These results indicate a relationship between superfluid density, local strain, and frustrated magnetism, and demonstrate two mechanisms for enhancing critical currents.

The iron arsenide superconductors have critical temperatures $T_c$ up to 57 K, multi-band Fermi surfaces, and undoped parent compounds that have an antiferromagnetic, orthorhombic state below temperatures of ~100-200 K. In Ba(Fe$_{1-x}$Co$_x$)$_2$As$_2$ and other members of the 122 family (AFe$_2$As$_2$ with A = Ca, Sr, Ba), doping causes the spin density wave transition temperature $T_{SDW}$ and the structural transition temperature $T_S$ to decrease [1,2], falling to zero at or near the doping where the highest $T_c$ occurs, suggesting the importance of lattice changes in determining transport properties. Proposals for magnetic mechanisms for superconductivity range from antiparamagnons [3] to magnetic antiphasons [4]. The existence of a large magnetoelastic effect in calculations [5], the close relationship between the structural and magnetic transitions [6], and the observation of a large and similar iron isotope effect for both $T_c$ and $T_{SDW}$ [7] all indicate the close relationship between structural and magnetic properties, to the point that some authors describe both the lattice and the SDW with a single order parameter [8,9]. Structural inhomogeneity is also important: in the 122 compounds, very small amounts of non-hydrostatic pressure can induce superconductivity [10-13], suggesting that structural inhomogeneity enhances superconductivity. Lattice changes due to pressure and chemical doping are remarkably similar [14,15]. The recent observation of twin boundaries in the 122 parent compounds [16] offers a controlled source of structural variation. We find that images of the local diamagnetic susceptibility (Figure 1) of single crystals of underdoped Ba(Fe$_{1-x}$Co$_x$)$_2$As$_2$ show a striped pattern for $T \le T_c$, and we show that these features are consistent with a large enhancement of the superfluid density on the twin boundaries.

The crystal growth is described elsewhere [1]. We use a variable-$T$ scanning SQUID susceptometer [17-19] with two field coil / pickup loop pairs (Figure 1a overlay sketch) in a gradiometer configuration. Each pair has a mutual inductance $\chi$ = 800 $\phi_0$/A. The susceptibility of the pair that is scanned close to the sample is modified by an amount $\Delta\chi$ [20]. The exact relationship of $\Delta\chi$ to the local magnetic penetration depth $\lambda$ and superfluid density $n_s \sim 1/\lambda^2$ is determined by the sample and sensor geometry as described in [17-21]. For a uniform sample in the limit of small $n_s$, $\Delta\chi \propto n_s$. The spatial resolution is primarily determined by the pickup loop diameter (3 μm), although the field coil diameter (13 μm) and sample-sensor separation (1-2 μm) also play a role.

We observe lines of enhanced diamagnetic susceptibility, henceforth called "stripes", along the orthorhombic [110] or [1$\bar{1}$0] crystalline directions. The spacing of the stripes is not periodic and varies between 10-16 μm in Figure 1, up to 35 μm in other parts of this sample, and down to 5 μm in parts of other samples. There were also regions of each sample that did not show any stripes.

The spacing and orientation of the stripes, the known orthorhombicity of Ba(Fe$_{1-x}$Co$_x$)$_2$As$_2$ on the underdoped side, and the recent demonstration of the existence of twin boundaries in the parent compound [16], all suggest identifying the stripes with twin boundaries (Figure 1d-g). Recent local polarization and x-ray measurements by Tanatar *et al.* [16] in the undoped parent compound BaFe$_2$As$_2$ have confirmed that the twin boundaries form in stripes spaced by around 10-50 μm, and shift position between thermal cycles above $T_S$.

We checked the hypothesis that the stripes are twin boundaries in three ways. First, we checked their behavior on thermal cycling (Figure 2). The stripes disappear on increasing $T$ above $T_c$, but reappear in the same locations when cooled as long as $T$ is not raised above $T_s$. However, each time $T$ is raised above $T_S$ and then decreased below $T_c$, the stripes reappear in somewhat different locations, consistent with twin boundaries, which would be formed as $T$ decreases through $T_S$.

Second, in all the samples studied (Table 1), the stripes appeared only in underdoped samples, consistent with twin boundaries, which would not exist in the tetragonal overdoped samples.

Third, we made $T$ dependent powder x-ray diffraction measurements on samples from x = 0% to 5.1%, which confirmed the existence of the tetragonal to orthorhombic phase transition at a doping dependent $T$ from 135K to 55K. We also confirmed the existence of a structural phase transition on three single-crystal samples with dopings of 0%, 2.5%, and 5.1% using spatially resolved single-crystal x-ray diffraction with a beam spot of 10 μm. We found that each sample formed twins with at least two different orientations, again consistent with the hypothesis of enhanced diamagnetic susceptibility on twin boundaries.

The susceptibility of each sample (Figure 3a) was measured as the difference between the $\Delta\chi$ high above the sample and in contact at a specific location (Method 1) or $\Delta\chi$ averaged over a region of the sample (Method 2). The difference in low-$T$ $\Delta\chi$ between UD1 and UD3 is likely due to uncertainty in the sensor-sample separation. As expected, $\Delta\chi$ decreases with increasing $T$ until it disappears at $T_c$, where $\lambda$ diverges. The locally measured $T_c$ typically varied by ~±0.25K for different regions of the samples with x = 5.1% (UD1, UD3, and UD5), by ~±1K for samples with x = 4.5% (UD2 and UD4), and by ~±0.1K for samples with x = 6.1% and 8.5% (OD1, OD2, and OPD1).

The magnitude of the susceptibility signal associated with the stripes, $\Delta\chi_{stripe}(T)$, for three individual stripes is shown in Figure 3b (from the stripe images on the right side of figure 4b and 4d). The typical $T$-dependence in both samples is similar (Figure 3c). $\Delta\chi_{stripe}$ ~ 2-6 $\Phi_0/A$ is comparable to the bulk susceptibility signal close to $T_c$, but much smaller than the saturated bulk susceptibility $\chi_{sample}$ ~ 600 $\Phi_0/A$ at lower $T$. The reduction of $\Delta\chi_{stripe}$ at lower $T$ likely reflects more effective screening from the bulk.

The spatial width of every stripe was apparently limited by our spatial resolution. To estimate the superfluid density associated with $\Delta\chi_{stripe}$, we used fasthenry (Whiteley Research) to model a stripe as a vertical slab of width $w$ between 0.2 and 5.0 μm with penetration depth $\lambda_{stripe}$ between 0.2 and 5 μm. For narrow vertical slabs with enhanced $n_s$, we found $\Delta\chi$ proportional to sheet density $n_sw$, with $\Delta\chi_{stripe}$ ~ 6 $\Phi_0/A$ corresponding to $n_sw$ ~$10^{19}$ m$^{-2}$, assuming a bulk penetration depth $\lambda_{bulk} \geq 10$ μm. Incorporating stronger Meissner screening in the bulk (with $\lambda_{bulk}$ between 0.2 and 10 μm) reduced the signal, implying that $10^{19}$ m$^{-2}$ is an order of magnitude lower bound.

The stripes dramatically affect the behavior of individual vortices. Figure 4 compares susceptometry with magnetometry images. The magnetometry images show individual vortices, pinned on local defects, whose small density is consistent with the small magnetic field in the cryostat. These images are similar to those observed in other type II superconductors, except that in multiple thermal cycles on multiple samples, every vortex avoids pinning directly on a stripe. The vortices also avoid the stripes when moving under the influence of temperature or the SQUID. Close to $T_c$, where vortices are weakly pinned and easily moved, they tend to move along the nearest stripe, and avoid crossing it. In figure 4g, we applied a small dc current to the field coil to create a controlled ~0.7 pN force [22] on the vortex shown while imaging it at 15.5 K: despite the applied force, it did not cross the stripe but moved easily along it.

As $T$ increases, the vortices spread out and become more mobile until at 18K we cannot detect them. At these temperatures the stripes can be seen in magnetometry images with a small amplitude of ~ 0.03 $\Phi_0$. We suspect that the reason the stripes are visible in magnetometry here is the distorted magnetic shape of the vortices, which, while very large, still terminate on the stripes.

That vortices avoid the stripes is consistent with enhanced superfluid density on the stripes: it is more energetically costly to create a vortex core in a region of enhanced superfluid density. This behavior is in sharp contrast to the behavior observed in $ErNi_2B_2C$, where, in a small magnetic field, all vortices are observed to sit specifically on twin boundaries in the antiferromagnetic and superconducting state [23-25]. The behavior of avoiding twin boundaries is also opposite to the familiar picture of twin boundaries in the cuprate superconductor YBCO, where vortices are either trapped on the boundary or channel along it, in both cases preferring to be co-located with the twin boundary [26-28].

All the data are consistent with a substantial enhancement of superfluid density on the twin boundaries in underdoped $Ba(Fe_{1-x}Co_x)_2As_2$, persisting to within at least 0.2 K of $T_C$. Two immediately apparent explanations for a variation in $n_s$ at a twin boundary are unlikely to apply here. First, variation in chemical doping across the twin boundary is unlikely, because the structural transition occurs well below 200 K. Second, in elemental

Sn or Nb, there is enhanced $T_c$ at a twin boundary due to phonon softening [29]. However, conventional electron-phonon coupling alone cannot explain the $T_c$ of the pnictides [30], and the dominant effect that we report here is an enhanced $n_s$.

More exotic categories of explanations include the possibility that the local symmetry of the twin boundary allows competing superconducting order parameters to arise, or that the presence of the twin boundary modifies the spin density wave and allows superconductivity to gain the upper hand in the competition between the two. In Figure 1g we have drawn the co-linear antiferromagnetic ordering of the Fe spins below $T_{SDW}$ in the configuration they would have without the twin boundary. It is clear that along the twin boundaries the spin orientation is ill-defined, which may enhance spin-fluctuation mediated pairing along the boundary.

The present measurements may be related to results that connect structural/magnetic deformations to superconductivity[10-12,31]. For example, the multi-phase structure present under non-hydrostatic pressure may have an increased number of planar structural inhomogeneities[10-12]. As another example, $SrFe_2As_2$ shows a decreasing volume fraction of superconductivity with increasing $T_c$ vs. pressure on the underdoped side of the phase diagram in [31]. Since moving along the pressure axis changes the surface area of twin boundaries as well as their strain, our results would provide a natural reason why $T_c$ might not follow the bulk superconducting transition.

The observation of enhanced superfluid density along two-dimensional planes, which likely are twin boundaries, indicates the need for caution in interpreting sample-averaged measurements in terms of homogeneous properties. In addition, it provides a physical realization of a relatively well-understood source of inhomogeneity that has a major effect on superfluid density. Any correct theory of the relationship between structure, superconductivity, and magnetism should be able to explain the presence of an enhanced superfluid density on the twin boundaries.

Finally, we note that the behavior we observe here should increase the bulk $J_c$ in two ways: by providing a channel for increased supercurrent density along the twins, and by providing strong barriers to vortex motion across the twins. These results suggest that increasing twin density should increase the sample-averaged bulk $J_c$.

**Acknowledgements**

We would like to thank O.M. Auslaender, Lan Luan, A. Auerbach and A. Vishwanath for useful discussions. This work was supported by the Department of Energy DOE Contract No. DE-AC02-76SF00515, by NSF Grant No. PHY-0425897, and by the US-Israel Binational Science Foundation. BK acknowledges the support of the L'Oreal USA Fellowships For Women in Science program. JRK was partially supported with a Humboldt Forschungspreis. Single crystal x-ray diffraction was performed at HPCAT (Sector 16), Advanced Photon Source (APS), Argonne National Laboratory. HPCAT is supported by DOE-BES, DOE-NNSA, NSF, and the W.M. Keck Foundation. APS is supported by DOE-BES, under Contract No. DE-AC02-06CH11357. Powder x-ray diffraction is carried out at Stanford Synchrotron Radiation Laboratory, a national user facility operated by Stanford University on behalf of the U.S. Department of Energy, Office of Basic Energy Sciences. We thank Wenge Yang for the support of the single crystal x-ray diffraction measurements, and A.S. Erickson, C.L. Condron, and M.F. Toney for their support of the powder x-ray diffraction experiments.

**Figure Captions**

**Figure 1:** Local susceptibility images in underdoped Ba(Fe$_{1-x}$Co$_x$)$_2$As$_2$, indicating increased superfluid density on twin boundaries. **(a)** Local diamagnetic susceptibility, at T=17 K, of the *ab* face of sample UD1 (x = 0.051, $T_c$ = 18.25 K), showing stripes of enhanced diamagnetic response (white). **Overlay:** sketch of the scanning SQUID's sensor. The size of the pickup loop sets the spatial resolution of the susceptibility images. **(b,c)** Images of the same region at T=(b) 17.5 K and (c) 18.5K show that the stripes disappear above $T_c$. **(d)** BaFe$_2$As$_2$ unit cell in the **(e)** tetragonal and **(f)** orthorhombic phases (orthorhombic distortion exaggerated for clarity). **(g)** Possible twin boundary configuration. Spins on the Fe sites are drawn in the configuration that they would have in the absence of the twin boundary.

**Figure 2:** Effect of thermal cycling on the locations of the stripes. **(a-c)** Local susceptibility images at 17 K (a) before and (b) after thermal cycling to T=25 K, above $T_c$ but below $T_{S/SDW}$, show (c) unchanged stripe locations. **(d-f)** Images (d) before and (e) after thermal cycling to 90 K, above $T_{S/SDW}$, show (f) changed stripe locations.

**Figure 3:** Temperature dependence of the local diamagnetism. **(a)** Typical susceptibility signal Δχ(T) in samples UD1 and UD3 vs. $T/T_c$. **(b)** Amplitude for three individual stripes in UD3 (stripes shown in Figs. 4b and 4d). **(c)** Average stripe amplitude for one region of UD1 and two different regions of UD3. T is scaled by the local $T_c$=18.5K for UD1 and 18K for UD3.

**Figure 4:** Interaction between vortices and stripes. **(a)** Magnetometry and **(b)** susceptometry images of the same region of UD3 at 5 K. Comparison of stripe (b) to vortex (a) locations shows that no vortex is pinned on a stripe. Light grey disks at the vortex locations in (b) are artifacts related to nonlinearity in the SQUID feedback loop at this set-point. **(c,d)** show the same location imaged at 15K, where the vortices are dragged by an attraction to the SQUID sensor, and tend to move along the stripes. **(e)** Overlay of susceptometry and magnetometry images in UD1 at 17.5K, where the vortices are wider. The vortices move during the scan, generally avoid crossing the stripes, and show a distorted shape near the stripes. Magnetometry **(f)** and susceptometry **(g)** images of a single vortex sitting between stripes in UD3, imaged with a current on the field coil to apply an attractive field coil - vortex force of ~0.7pN. The vortex is dragged by the field coil at 15K, but only moves between the stripes, not across them. Arrows in (f) show scan directions: the sensor is incremented along the long arrow while rastered back and forth along the double arrows.

**Table 1. Samples**

| Name | Doping | $T_C$ | Measurement |
|---|---|---|---|
| UD1 | 5.1% | 18.25±0.25K* | SSM |
| UD2 | 4.5% | 12.75±0.5K* | SSM |
| UD3 | 5.1% | 18.25±0.25K* | SSM |
| UD4 | 4.5% | 12.25±1K* | SSM |
| OD1 | 8.5% | 19.9±0.1K* | SSM |
| OD2 | 8.5% | 20.2±0.1K* | SSM |
| OPD1 | 6.1% | 22.8±0.1K* | SSM |
| ND1 | 0% | - | XRD |
| UD5 | 2.5% | - | XRD |
| UD6 | 5.1% | 18.7K±0.25K** | XRD |

*$T_C$ reported is measured in situ.

**$T_C$ of sample batch measured by susceptibility using Quantum Design 5T MPMS.

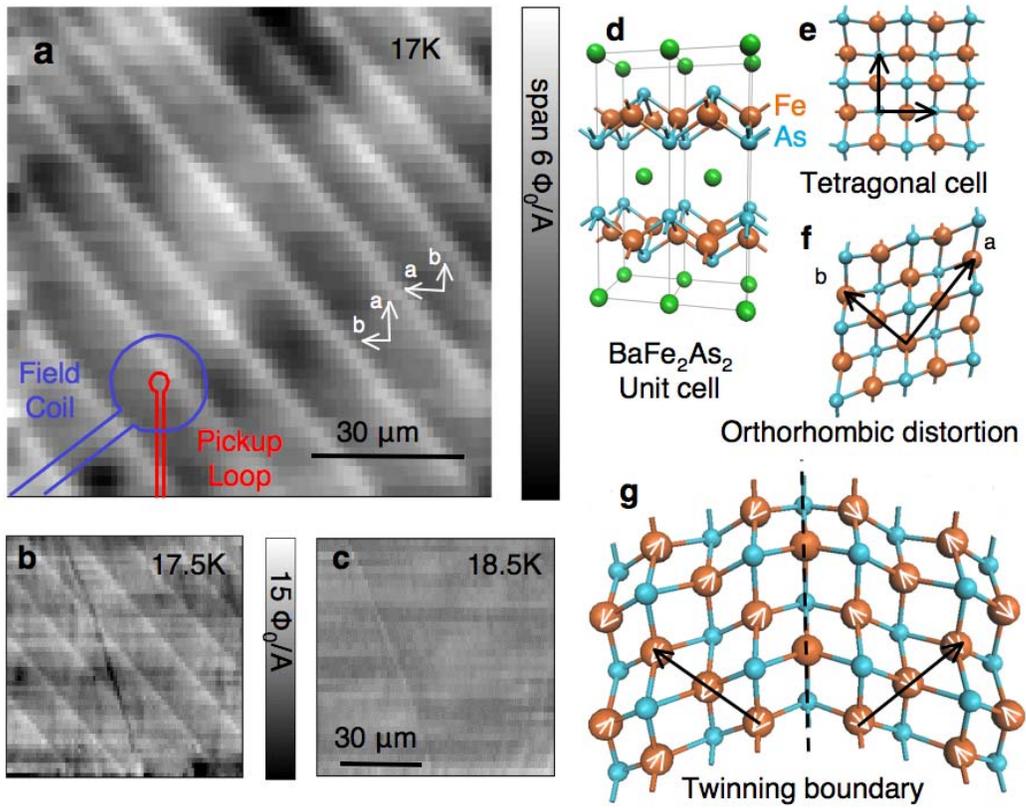

**Figure 1.**

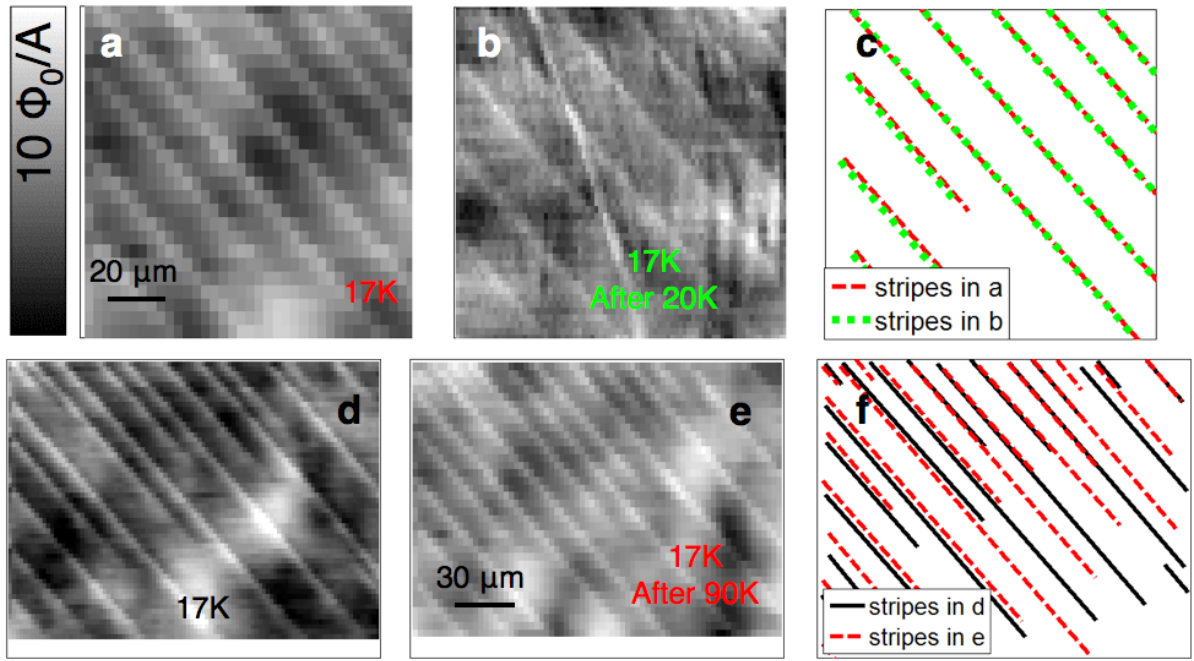

**Figure 2.**

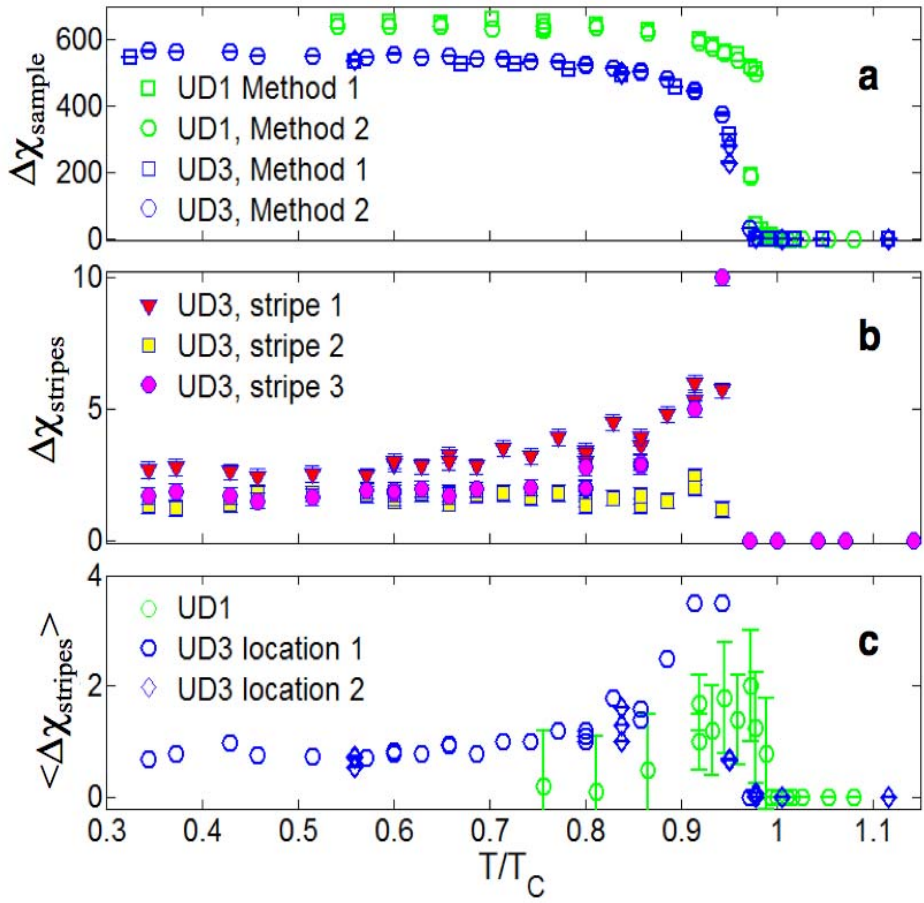

**Figure 3.**

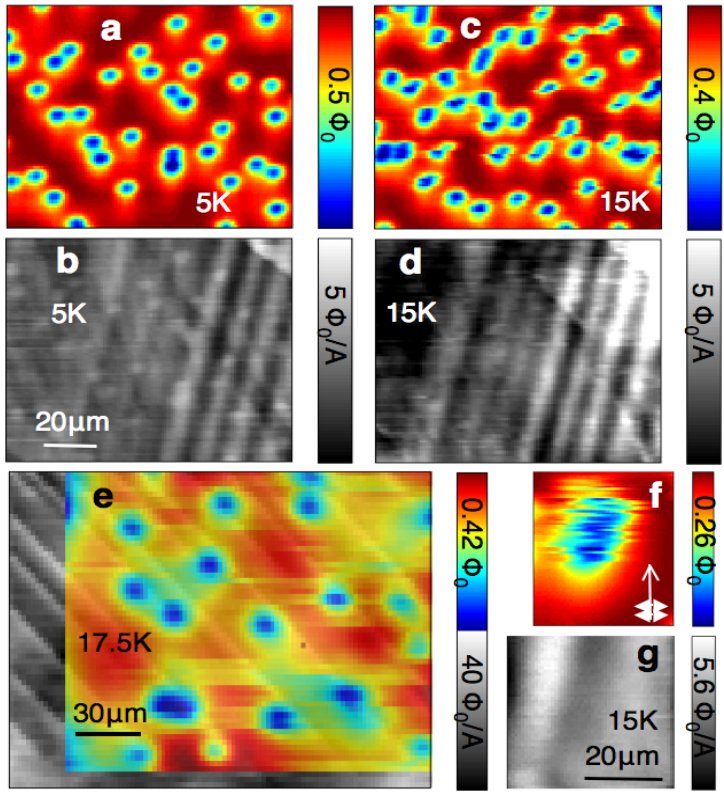

**Figure 4.**